\documentclass[showpacs,preprintnumbers,amsmath,amssymb]{revtex4}

\usepackage{pstricks}
\usepackage{graphicx}
\usepackage{dcolumn}
\usepackage{bm}

\begin{document}

\newcommand{\locsection}[1]{\setcounter{equation}{0}\section{#1}}
\renewcommand{\theequation}{\thesection.\arabic{equation}}

\def\F{{\bf F}}
\def\A{{\bf A}}
\def\J{{\bf J}}
\def\af{{\bf \alpha}}
\def\beqn{\begin{eqnarray}}
\def\eeqn{\end{eqnarray}}

\def\dspace{\baselineskip = .30in}
\def\beq{\begin{equation}}
\def\eeq{\end{equation}}
\def\bw{\begin{widetext}}
\def\ew{\end{widetext}}
\def\pl{\partial}
\def\na{\nabla}
\def\al{\alpha}
\def\bt{\beta}
\def\Ga{\Gamma}
\def\ga{\gamma}
\def\de{\delta}
\def\De{\Delta}
\def\da{\dagger}
\def\ka{\kappa}
\def\si{\sigma}
\def\Si{\Sigma}
\def\te{\theta}
\def\La{\Lambda}
\def\lam{\lambda}
\def\Om{\Omega}
\def\om{\omega}
\def\ep{\epsilon}
\def\non{\nonumber}
\def\sq{\sqrt}
\def\sqg{\sqrt{G}}
\def\sp{\supset}
\def\sb{\subset}
\def\l{\left (}
\def\r{\right )}
\def\lq{\left [}
\def\rq{\right ]}
\def\fr{\frac}
\def\la{\label}
\def\hs{\hspace}
\def\vs{\vspace}
\def\inf{\infty}
\def\ran{\rangle}
\def\lan{\langle}
\def\ov{\overline}
\def\tl{\tilde}
\def\tm{\times}
\def\lrar{\leftrightarrow}




\vs{1cm}

\title{Trinification Phenomenology \\
and the structure of Higgs Bosons}

\author{Berthold Stech}
\email{B.Stech@ThPhys.Uni-Heidelberg.DE}

\affiliation{Institut f\"ur Theoretische Physik, Philosophenweg 16, D-69120 Heidelberg, Germany}

\vs{1cm}


\begin{abstract}

The extension of the Standard Model to $SU(3)_L \times SU(3)_R \times SU(3)_C$ 
(the trinification group)  augmented by  the $ SO(3)_G $ flavor group is considered.  In our phenomenological treatment partly known and partly proposed vacuum expectation values of the scalar Higgs fields play a dominant role.  All Higgs fields are taken to be flavor singlets, all flavon fields trinification singlets. We need two flavor (generation) matrices. One determines the mass hierarchy of all fermions, the second one is responsible for all mixings including the CP-violating phase in the CKM matrix. The mixing with higher states contained in the group representation provides for an understanding of the difference between the up quark and the down quark spectrum. There is a close connection between charged and neutral fermions.  An inverted neutrino hierarchy is predicted. Examples for the tree-level potential of the Higgs fields are given. To obtain an acceptable spectrum of scalar states, the construction of the potential requires the combination of matrix fields that differ with respect to fermion couplings and flavor-changing properties. As a consequence bosons with fermiophobic components or, alternatively, flavor-changing components are predicted in this model. Nevertheless, the Higgs boson at $125~$GeV is  very little different from the Standard Model Higgs boson in its couplings to fermions but may have self-coupling constants larger by a factor 2. 

\end{abstract}

\maketitle

\section{The Model}\label{sec:1}

According to present experimental results at the Large Hadron Collider supersymmetry has not been observed \cite{Craig}. It may not be relevant at the weak scale. The hierarchy problem still persists. This is a serious problem for the vacuum expectation values (vevs) of scalar fields. These important momentum independent (apart from wave function renormalization) quantities are presently not understood. They may have their origin at a very high scale. In the Standard Model  the  vev of the Higgs field is the cause of all particle masses. Also the well-known quadratic divergence of the Higgs self-energy caused by fermion loops is related to this vev. Thus the vacuum expectation value can be viewed as the physics origin of the Higgs and fermion masses. In this article we try to keep this feature by dealing with the vevs of extended models. 

We treat all vacuum expectation values of scalar fields as fundamental fine-tuned parameters.  In addition, only  dimensionless coupling constants are used.  
 All masses are obtained in terms of these vevs and these dimensionless coupling constants. The idea is to start with a massless Lagrangian. By introducing vevs  for the scalar fields,  linear field components  show up which have to be canceled. In a simple  $ \phi^4 $ model this cancellation is performed by adding a term $\phi^2$  multiplied by the square of the corresponding vev:
\begin{eqnarray}
\label{1.1}
\lambda~\phi^4  \rightarrow \lambda ~\phi^4 - 2 \lambda ~\langle \phi \rangle ^2 ~\phi^2 .
\end{eqnarray} 

Now the shift $ \phi \rightarrow   \langle \phi \rangle + ~ \phi' $ ~can be applied and no linear term in $\phi'$ appears anymore. The second  derivative of this expression provides the Higgs mass in terms of the vev $\langle \phi \rangle$ : $m^2_H= 4 \lambda \langle \phi \rangle ^2$ .  

If log terms can be used more possibilities are open. For instance the modified form
\begin{eqnarray}
\label{1.2}
\lambda~\phi^4  \rightarrow \frac {\lambda}{1+\log[\frac{\phi^4}{ \langle \phi \rangle ^4}]} ~ \phi^4
\end{eqnarray} 
has no linear term in $\phi'$ either. The mass at the minimum of this potential is $ m^2_H = 8  \lambda \langle \phi \rangle ^2 $.

The embedding of  the Standard Model into a larger group allows us to connect  the properties of quarks and charged leptons and their mixings with the properties of neutrinos and their different mixings. It also implies a Higgs sector with more scalar bosons.  Here we extend the Standard Model symmetry group $ SU(2)_L \times U(1) \times SU(3)_C$ to $SU(3)_L \times SU(3)_R \times SU(3)_C $, the trinification group \cite{Trini1,Trini2, TriniPak}, which is a subgroup of $E_6$ \cite{E61,E62,E63,E64}. 
In this report we are guided by articles using $E_6$ \cite{BZ1,BZ2,BZ3} but restrict ourselves to the simpler trinification subgroup. See-saw formulae for quarks and leptons given there can be used. Some details are different and numerical updates are performed. Two Higgs fields with antisymmetric flavor couplings need special attention. We treat the fermion and the scalar sectors and discuss their connections.~The model contains a large number of scalar fields. In spite of these numbers tree-level potentials providing phenomenologically acceptable mass values can be obtained which is hardly possible in a full $E_6$ model. Suggestions  in \cite{BHiggs1, BHiggs2, BTHiggs} are used and explored. But at present only examples for the scalar particle spectrum and state mixings can be given. However, they show highly interesting qualitative properties: almost all scalar states have fermiophobic or flavor-changing components.

The group $SU(3)_L \times SU(3)_R \times SU(3)_C $  can be unbroken only at and above the
scale where the two electroweak gauge couplings $g_1$ and $g_2$ combine. According to the scale dependence of the Standard Model couplings this happens at a scale of about $10^{13}$ to $10^{14}~{\rm GeV}$. Interestingly, this is just the scale relevant for the small values of the neutrino masses by applying the seesaw mechanism. It is also the place where the self coupling of the Higgs field approaches  zero \cite{zero}.  In this article we do not consider the possible complete unification of $g_1$, $g_2 $ and  $g_3$ expected at  a still higher scale.

All fermions are described by 
two-component  (left-handed) Weyl fields. As abstracted from the $\bold 27$ representation of $E_6$ \cite{BZ1,BZ2}.
they occur in singlet and triplet $SU(3)$ representations of the
trinification group with the quantum number assignments 
\beqn
\label{fermions}
{\rm Quarks}:~~~q(x)=(3, 1, \bar 3) ,
\non
\\
{\rm Leptons}:~~~L(x)=(\bar 3, 3, 1) ,
\non
\\
{\rm Antiquarks}:~~~\hat q(x)=(1, \bar 3, 3) .
\la{ab}
\eeqn
For each generation one has
\beqn
\begin{array}{cc}
 & {\begin{array}{cc}
 \hspace{1.8cm}&
\end{array}}\\ \vspace{2mm}
q_i^a=\hs{-0.2cm}
\begin{array}{c}
 \\

\end{array}\!\!\!\! &{\left(\begin{array}{ccc}
\hs{-0.3cm}u^a
\\
\hs{-0.2cm}d^a
\\
\hs{-0.2cm}D^a
\end{array}\hs{-0.2cm}\right)\! },\hspace{1.5cm}
\end{array}
\hs{-0.1cm}
\begin{array}{ccc}
& {\begin{array}{ccc}
 & &
\end{array}}\\ \vspace{2mm}
~~~~~~~~^i L_k= \hs{-0.5cm}
\begin{array}{c}
  \\
\end{array} \hspace{-0.1cm}&{\left(\begin{array}{ccc}
^1L_1 & E^{-} & e^{-}
\\
E^{+} & ^2L_2 & \nu
\\
e^{+} & \hat{\nu } & ^3L_3
\end{array}\right),
}
\end{array} ~~~~~~~
\hat q^k_a=\l \hat{u}_a,~\hat{d}_a,~\hat{D}_a \r ,\hs{1.3cm}
\label {ac}
\eeqn
where $i, k, a=1, 2, 3$.
In this description $SU(3)_L$ acts vertically (index $i$) and $SU(3)_R$
horizontally (index $k$) and  $a$ is a color index. 

As proposed in the articles quoted earlier  we use also the generation (flavor) group $SO(3)_G$ and require all fermions to be $3$-vectors in generation space. The coupling matrices for fermions are vevs of flavon fields. In our phenomenological treatment these couplings are considered as parameters of the model, without regarding the flavon potential from which they originate. We  take all Higgs fields to be singlets with respect to the flavor symmetry and all flavon fields to be singlets with respect to the trinification group.  

 Generation (flavor) indices will be denoted by $\alpha,\beta =1,2,3$. The coupling matrices occurring in the Yukawa interaction originating from trinification singlet flavon fields are taken to be hermitian $ 3 \times 3 $ matrices. Two coupling matrices are needed: The symmetric matrix  $ G_{\alpha,\beta}$ and the antisymmetric matrix $A_{ \alpha,\beta}$.   By a $SO(3)$ symmetry redefinition $ G_{\alpha,\beta}$  can be taken to be a diagonal matrix. It  is responsible for the flavor hierarchy of all fermion masses.  The antisymmetric matrix $A_{ \alpha,\beta}$ determines the fermion  mixings.

The scalar bosons (Higgs fields) are described by the matrix fields $H$, $H_A$, $H_{A l}$ and $\tl H$. 
They transform under the trinification group as $(\bar 3, 3, 1)$ except for $H_{A l}$ which transforms according to $(\bar 3, \bar 6, 1)$.
 These matrix fields differ with respect to their Yukawa couplings: $H$ is coupled to fermions with the symmetric coupling matrix $ G_{\alpha,\beta}$ whereas  $H_{A}$ and $H_{A l}$ couple to fermions with the antisymmetric flavor matrix $A_{ \alpha,\beta}$. As deduced from $E_6$,  $H_{A }$ acts on quarks, $H_{A l}$ on leptons only  \cite{BZ1,BZ2}.
$\tl H $, on the other hand, does not couple to fermions . The latter property can be achieved by an additional parity-like symmetry $P_G$ with a positive value for 
 $H$, $H_A$, $H_{A l}$ and a negative value for the fermiophobic $\tilde H$. In order to obtain a spectrum of  the scalar fields with non-vanishing masses (besides the would-be-Goldstone states)  the potential has to be formed from at least two different matrix fields. 
This fact leads us to consider two distinguished cases:

A) The fermiophobe model: the potential is formed by $H$ and $\tilde H$ while the fields in $H_A$, $H_{A,l}$ are assumed to have a negligible influence on the lower part of the mass spectrum of the scalars. In this case most states will have fermiophobic components.

B) The flavophile model: Here we take $\tilde H=0$ and construct the potential out of the fields  $H$ and $H_{A}$. The contribution of $H_{A l}$ to the potential is assumed not to be relevant for the low mass eigenstates, or $H_{A l}$ has its own additional potential independent of $H$. In this case the scalar bosons can cause flavor-changing transitions as we will see.

All scalar fields that are neutral and color singlets can in general have vaccuum expectation values with different magnitudes.  $\langle H \rangle $ as well as  $\langle \tilde H \rangle $ have then the form of the lepton matrix (\ref{ac}) with the charged components set to zero i.e. five non-zero elements. Biunitary $SU(3)_L$ and $SU(3)_R$ $U$-spin transformations can then be applied to bring the 
matrix $\langle H \rangle $ to a diagonal form. This defines $D$, $\hat D $ as well  as the other fermions in (\ref{ac}) to be  eigenstates of  $\langle H \rangle $ after symmetry breaking.
With the  vev of $H$ being a diagonal matrix the vev of  $\tilde H$ will  in general have $5$ non-zero elements.
\beq
\label{vevH}
\langle H \rangle \hs{-0.8mm}=\hs{-0.8mm}
\left(\begin{array}{lll}
v_1&~~0&~0\\
0&~~b &~0\\
0&\ 0 & M_I \end{array}
\right),~~
\lan \tl H \ran \hs{-0.8mm}=\hs{-0.8mm}
\left(\begin{array}{lll}
v_2&0&0\\
0& b_2 & b_3\\
0& M_R & M_3
\end{array}
\right).
\end{equation}

 The vev of the matrix field $H_{A} $ also has $5$  non-zero elements  in general. We write them in the form $^i f_k$.  Here $i$ is again a left-handed  index and  $k $ a right-handed one with $i, k=1, 2, 3$. The matrix field $H_{A l}$ can have $8$ non zero elements occurring in the expression
  $^i f^{\{k l\}}$.   (For brevity these indices will often be omitted hereafter).
 A number of these vevs for  $H_A$ and $H_{A l}$ are taken to be zero since all mixings are supposed to originate from the mixings of Standard Model fermions with the high-mass fermions present in the representation (\ref{fermions}). Since the top has no higher partner we  take $^1f_k =~^1f^{\{k l\}} = 0$.  
$M_I$ together with $M_R$ and/or $^3f_2$  
 break the trinification group down to the Glashow-Weinberg-Salam group. They have to be large compared to the weak scale.  $M_I$ is presumably close to the meeting point of $g_1$ and $g_2$ mentioned above.  The vevs in the first and second rows are assumed to be of the order of the weak scale or smaller. They break the Standard Model group down to the electromagnetic $U_e(1)$ symmetry. 
As a  consequence of our scheme it is seen that $v_1$ and $b$ are related to the top mass and the (unmixed) bottom quark mass:
\beq
\label{t,b}
m_t = g_t ~ v_1,~~ m^0_b = g_t~b ~~~~~i.e.~~  \frac{b}{v_1}=\frac{m^0_b}{m_t} .
\eeq
The known value $v=174$~GeV ($= \frac {246}{\sqrt{2}} $~GeV) for the vev of the Higgs 
field  of lowest mass is related to  vevs of $H$, $\tl H$,  $H_A$ and $H_{A l}$ :
\beq
\label{v}
v^2 = v_1^2 + v_2^2 +b^2+b_2^2+b_3^2 + (^2f_2)^2 +(^2f_3)^2  + (^2f^{\{1, 3\}})^2 + (^2f^{\{1, 2\}})^2.
\eeq
Thus, if the  $b_j$ and $f_j$ are similarly small as $b$, ($b\approx m_b \simeq 2.85~\rm GeV$ at the scale $m_Z$) the vevs  $v_1$ and $v_2$ are restricted according to $v_1^2 + v_2^2 \simeq v^2$.

With the flavor matrices $G$ and $A$ and the flavor singlet Higgs fields $H$, $H_A$ and $H_{A l}$ the Yukawa interaction  \cite{BZ2,BZ3} is
\beqn
\label{LY}
{\cal L}^{1}_Y~ = ~g_t G_{\alpha \beta} 
\left(\psi^{\al T} H~\psi^{\beta} \right)+   A_{\al \beta} 
\left(\psi^{\al T} H_{A}~\psi^{\bt}\right) +   A_{\al \beta} 
\left(\psi^{\al T} H_{A l}~\psi^{\bt}\right) ~+~ h.c.
\eeqn
Here $\psi$ stands for a column vector consisting of all fermions described in (\ref{fermions}). But as mentioned before, $H_A$ acts only on quarks and $H_{A l}$  on leptons only.
 The first term gives  the up quarks their masses. It also describes parts of the down quark and lepton mass matrices (with Dirac masses for the neutrinos). The second term in (\ref{LY}) performs the mixings of the Standard Model quarks as well as their mixings with the heavier quark states occurring in the model.
Together,  the first and the second term should achieve satisfactory results for the masses of up quarks,  down quarks and the Cabibbo-Kobayashi-Maskawa matrix including its CP-violating phase. Similarly, the first term together with the third term should describe masses and mixings of the leptons.

 At this stage, however, the neutrinos are still Dirac neutrinos with masses comparable to the quark masses. Moreover,  neutral leptons that are singlets with respect to Standard Model gauge transformations are still massless. An important assumption of our model is therefore the addition of an effective Yukawa interaction
  ${\cal L}^{\rm eff,2}$.  
\beqn
 \label{LY2}
{\cal L}^{\rm eff,2} =  \frac{1}{M_N} \left( G^2 \right)_{\al \bt} 
\left((\psi^{\al T}  H^{\dagger})_1(\tilde H^{\dagger} \psi^{\bt})_1 \right) +~h.c.~~ \text{or}~~~
{\cal L}^{\rm eff,2} = \frac{1}{M_N} \left( G^2 \right)_{\al \bt}  
\left((\psi^{\al T}  H^{\dagger})_1(H_A^{\dagger}~\psi^{\bt})_1 \right) +~h.c.
\eeqn
 
 As indicated by the index 1  (\ref{LY2}) couples fermions with $ H^{\dagger}$ and  similarly with $\tilde H^{\dagger}$ and  $H^{\dagger}_A$ to form trinification singlets . Since $H$ , $\tilde H$ and $H_A$ are matrices with left and right-handed indices this coupling clearly involves leptons only.  It could originate from the exchange of a very heavy trinification singlet neutrino with mass $M_N \approx M_I $ and an appropriate $U(1)_G$ structure. The first form is relevant for the fermiophobe model, the second is taken for the flavophile model 
 where $\tilde H = 0$. Here $\tilde H$ is replaced by $H_A$.  Obviously, (\ref{LY2}) provides masses for neutral leptons only. In particular, they provide a mass matrix with large eigenvalues (because of the large vevs $M_I$, $M_R$, $^3f_2$) to the neutrinos $\hat{\nu}=~^3L_2$ and $^3L_3$ which are not part of the Standard Model. Through the mixing of these heavy neutrinos with the Standard Model neutrinos the seesaw mechanism takes place. The corresponding flavor matrix in (\ref{LY2}) must be symmetric. Because of  the second order form of (\ref{LY2}) we choose the matrix $G^2$. As a consequence, the mass hierarchy of the heavy neutrinos is a very strong one \cite{BZ2,BZ3}. Thus, depending on the masses $M_N$ and $M_I$ the first generation of the leptons $\hat{\nu} =~ ^3L_2$ and $^3L_3$ can have masses in or below the TeV region.  

The effective Yukawa interaction ${\cal L}^{\rm eff} ={\cal L}^{\rm 1} +{\cal L}^{\rm eff,2}$ with $G$, $A$ in (\ref{LY}, \ref{LY2}) and the vev configurations of $H$, $H_A$, $H_{A l}$ and $\tilde H$ contain all  the necessary information about the generation structure and  the fermion spectrum. We will show in section 2 that the properties of quarks and charged leptons, their masses and mixings are well described by only a few parameters. The flavor matrices $G$ and $A$ determined in this section are then used in section 3 for the determination of the mass matrix for neutrinos. A generalized seesaw mechanism leads to an inverted hierarchy. Finally, in section 4,  tree-level potentials are constructed which give rise to spontaneous symmetry breaking and determine the mass spectrum of the Higgs-like bosons. The corresponding mass eigenstates of the scalar fields will in general have either fermiophobic or flavor-changing components.

\section{Charged Fermion Masses and Mixings}\label{sec:2}

\subsection{The up quark mass matrix}\label{a}

The experimentally known up quark masses can be used to construct the flavor matrix G. All experimentally determined masses used here and in the following are running masses taken at 
the scale $m_Z$. The numerical values based on recent determinations have been provided by Matthias Jamin \cite{Jamin} and are presented in the appendix.
 
 Since $G$ and $\langle H \rangle $ can be chosen to be diagonal matrices, one has from (\ref{LY}) 
  \begin{equation}
\label {G}
g_t v_1 G =\hs{-0.1cm}\left(\begin{array}{lll}
m_u&0&0\\
0&m_c&0\\
0&0&m_t
\end{array}\right)~
\end{equation}
By introducing the small parameter $\sigma= \sqrt{m_u/m_c}=0.045 $, a good phenomenological ansatz for $G_u$ for the  up quark masses at the weak scale 
is
  \begin{equation}
\label {G}
G_u=\hs{-0.1cm}\left(\begin{array}{lll}
p_t \sigma^4&0&0\\
0& p_t  \sigma^2&0\\
0&0&1
\end{array}\right)~.
\end{equation}
It contains the factor $p_t$. Interpreting
 this factor as a renormalization factor \cite{BZ2}  suggests that at a very high  scale one has $m_u/m_c= m_c/m_t = \sigma^2$. 
The up quark masses are well represented setting $p_t= 1.8$. In the next sections we need the matrix $G_d$ for the down quarks and $G_l$ for the charged leptons. For the down quarks the factor $p_t$ is replaced by $p_t^{1/3} =1.22 $  and for the charged leptons  by $1 $.  The small parameter $\sigma = 0.045$ introduced here can also be used to describe the particle mixing matrix $A$ and the neutrino properties.
 
 \subsection{The down quark masses and the Cabibbo-Kobayashi-Maskawa matrix}\label{b} 
 
The first two parts of the effective Yukawa interaction determine the down quark mass matrix. Here one needs  the antisymmetric flavor matrix $A$.  It can be described by $3$ real parameters.  We choose the element $A^1_2= i \sigma$ and absorb the remaining multiplication factor into the vev parameter of $H_A$. This leaves $2$ parameters for $A$. But in the form for $A$ that we will use here, only one real parameter ($\tau$) appears:

\begin{equation}
\label{A}
 A= ~i~\left(
\begin{array}{ccc}
0&\sigma&-\sigma\\
-\sigma&0&\tau\\
\sigma&-\tau&0
\end{array}\right)~.
\end{equation}
If  $\sigma / \tau$ would be equal to $1$, even (odd) permutations of generations would lead to 
$A\rightarrow +A$ ($A\rightarrow -A$).  The permutation symmetry with respect to the second and third generation survives for any value of $\tau$.
By fitting the CKM matrix the value $\tau = 0.50$ turns out to be a good choice.

Besides the down quarks of the Standard Model there exists - according to (\ref{fermions}) - also a state which is a singlet with respect to Standard Model gauge group transformations. Thus, the down quark mass  matrix is a $6\times  6$ matrix. This new quark $D$ with $SU(3)_L$ index $i=3$ is very heavy due to the vev $M_I$ of $H$.   One can integrate out this heavy state if the contributions from $H_A$ are taken to be small compared to $M_I$. This way one finds the wanted $3\times 3$ mass matrix for the Standard Model particles. The mixings with the high mass state cannot be neglected. It is seen to be essential for our understanding of the CKM matrix and the deviations of the mass pattern of down quarks from the mass pattern of the up quarks. The light down quark mass matrix is 

\beq
\label{md}
m_d= m^0_b~ G_d+ f^d  A +
f^d_0~\sigma^3  A ~(G_d)^{-1} A~.
\eeq

Here  $m_b^0= g_t  b $ is the value of $m_b$ before mixing, $f^d$ is equal to the vev 
of $^2(H_A)_2$ and $f^d_0$ a parameter resulting from integrating out the heavy $D$-quark masses. 
The factor $\sigma^3$ serves to cancel the negative powers of $\sigma$ in $ A (G_d)^{-1} A$ and thus
allows a smooth formal limit $\sigma \rightarrow 0$.  
Taking $m^0_b = 2.78$~GeV, $f^d = -0.280$ GeV and $f^d_0= 1.40 $ GeV  
(together with the values for $m_t$, $\sigma$ and $\tau$) an almost perfect representation 
for all up and down-quark properties is achieved. 
The masses at the scale $m_Z$ agree 
within error limits with the experimental ones \cite{Jamin}. 
Also the calculated CKM elements describe the data \cite{Fogli} quite well. The angles of the unitarity triangle come out to be
$\alpha \simeq 94^o, \beta \simeq 22^o, \gamma \simeq 64^o$. 

\subsection{The charged lepton masses and their (not directly observable) mixings}\label{c}  

  Like the down quarks the charged leptons have heavy partners as well and mix with them via the flavor 
  matrix $A$. Again, by integrating out these heavy states the $6 \times 6$ mass matrix is reduced to the 
  $3\times 3 $ matrix for the usual leptons. Apart from a sign, its form is the same as for the down quarks.
  \beq
\label{me}
m_e= - m^0_{\tau}~ G_l -  f^e  A   - f^e_0 ~\sigma^3 A ~(G_l)^{-1} A~.
\eeq
Here  $m_{\tau}^0$ is the value of $m_{\tau}$ before mixing. $m^0_{\tau}$, $f^e$ and $f^e_0$ are now used to fit the masses of $\tau$, $\mu$ and the electron. A good fit is obtained by setting $m^0_{\tau} = 1. 62$ GeV, $f^e = -0.2082$  GeV,
$f^e_0 = 2.58$  GeV. These values are not simply related  to the corresponding values for the down quarks since for leptons matrix elements of $H_{A l}$ instead of $H_A$ have to be taken \cite{BZ1}. The charged lepton mass matrix obtained this way allows to calculate the charged lepton mixings, which is a necessary ingredient for the discussion of the neutrino properties.

\section{Neutrino Masses and Mixings.}\label{sec:3}

According to the lepton assignments in (\ref{fermions}) 
 one has to deal with $5$ neutral leptons in each generation. Thus, the matrix for neutral leptons is a $15\tm 15$ matrix. Again, leptons that obtain high masses because of the large vevs  in the effective Yukawa interaction (\ref{LY}, \ref{LY2}) can be integrated out giving rise to a generalized seesaw mechanism. According to our Yukawa interaction the Dirac matrix for the light neutrinos is $ m_t G_l $, while the heavy neutrinos have masses proportional to $G_l^2$. Therefore, a unit matrix is part of the light neutrino mass matrix. Together with a further contribution due to the particle mixing matrix $A$
  one obtains  the ~$3 \times 3$ matrix  for the light neutrinos \cite{BZ2,BZ3}  in terms of the  two parameters $ \kappa$ and $m_0$ 
\beqn
\label{mnu0}
m_\nu \simeq \frac{m_t^2}{M_I } \kappa~ {\mathbf {1} }+ m_0 ~\sigma^3~ (A~ \frac{1}{G_l} - \frac{1}{G_l} A) .
\eeqn
Taking the neutrino mass matrix $m_\nu$ only up to first order in the small parameter $\sigma$, one can write $m_\nu$ in a very simple form:  setting 
$\kappa \frac{m_t^2}{M_I} = m_0~\rho$~ one obtains
\beq
\label{mnu}
 {\begin{array}{ccc}
 & &
\end{array}}\\ \vspace{1mm}
m_{\nu }\simeq 
  m_0~{\left(\begin{array}{ccc}
\rho &-{\rm i} & {\rm i}
\\
-{\rm i}  &\rho   & -{\rm i} \tau \si
 \\
 {\rm i}   & -{\rm i}\tau \si& \rho
\end{array}\right)}~
\eeq
with $\tau=0.50$ and $\sigma=0.045$ as used for quarks and charged leptons. The eigenvalues of $m_{\nu}.m^{\dagger}_{\nu}$ to first order in $\sigma $ are
\beqn
(m_2)^2 \simeq (\rho^2 +2 + \sqrt{2} ~\tau~\sigma)~ m_0^2~,
\non
\\
(m_1)^2 \simeq (\rho^2 +2 - \sqrt{2} ~\tau~ \sigma)~ m_0^2~,
\non
\\
(m_3)^2 \simeq \rho^2 m_0^2~.\hs{1cm}~
\label{eival}
\eeqn
It is now easy to see the following properties of the light neutrinos:

\begin{itemize}
\item  The neutrino mass spectrum has the form of an inverted hierarchy.
\item The ratio between the solar and the atmospheric mass squared differences is independent of the two neutrino parameters $m_0$ and $\rho$. This ratio is $\sqrt{2}~\tau~\si =  0.031$ in good agreement with experiment.
\item The experimentally observed atmospheric mass squared difference can be used to fix the mass parameter $ m_0$ for the light neutrinos.  Then $\rho$ is determined by the lightest neutrino mass $m_3$.
\beqn
\label{m3}
\vs{0.8cm}
 m_0 \simeq \frac{1}{\sqrt{2}} \sqrt{\Delta m^2_{\rm atm}}~ \simeq~ 0.035~{\rm eV}~,\hs{1cm} \nonumber  \\
 m_3 ~\simeq ~m_0~\rho ~ \simeq~ 0.035 ~\rho~ {\rm eV}.~~~~~~~~~~
\eeqn
\item Without taking account of  effects from diagonalizing the charged lepton mass matrix and renormalization, the neutrino mass matrix $m_\nu$ leads to almost strict  bimaximal mixing. 
\end{itemize}
Including the charged lepton mixings obtained from (\ref{me}) a better, but still not satisfactory agreement with the experimentally determined neutrino mixing angles is achieved. Detailed renormalization group calculations would be necessary, but are not performed  here. Instead we introduce a parameter which may in part simulate these effects. It should not change the successful mass pattern obtained so far and is therefore taken to be  
 an orthogonal transformation in generation space.  Mixing the first with the third generation by the angle $\phi$ one gets with  $c= cos~{\phi}$, $s=sin~{\phi}$ the modified neutrino mass matrix:
\beq
\label{mnuf}
 {\begin{array}{ccc}
 & &
\end{array}}\\ \hs{-0.5cm}
m_{\nu }\Rightarrow 
  m_0~{\left(\begin{array}{ccc}
\rho +2 {\rm i} c s &-{\rm i}(c+\tau \si s) & {\rm i} (c^2-s^2)
\\
-{\rm i}(c+\tau \si s)  &\rho   & -{\rm i} (\tau \si c -s)
 \\
{\rm i} (c^2-s^2) & -{\rm i} (\tau \si c -s)& \rho -2 {\rm i} c s
\end{array}\right)}.
\eeq
 Since $m_0$, $\tau$ and $\sigma$ are fixed, this matrix depends, for a given mass of  the lightest neutrino, only on the angle $\phi$. 
Of course, the mixing matrix of charged leptons,  obtainable from section $2$, has yet to be included.

As an illustrative example we take $\rho=1$ and choose  $\phi $  to fit  the third neutrino mixing angle
 $\te_{13}^{\nu } \approx 9^o$ in accord with data analysis \cite{Fogli}. This leads to $\phi \simeq \pi/14 $. One then finds for the neutrino masses and angles:
 \begin{eqnarray}
m_2=0.06012~ {\rm eV},~m_1=0.05948~ {\rm eV},~m_3=0.03456~ {\rm eV}
\nonumber  \\
\te_{12}^{\nu }\simeq 36^o, ~~~\te_{23}^{\nu }\simeq 49^o,~~~\te_{13}^{\nu }\simeq 9.4^o .~~~~~~~~~~~~
\hs{-0.3cm} 
\label{exa}
\end{eqnarray}
The CP-violating phase $\delta$, the Majorana angles and the mass parameter for the neutrinoless double $\beta $-decay in this example are
\begin{eqnarray}
 \de \simeq -11^o, ~
\frac{ \alpha_{2 1} }{2} \simeq -81^o, ~\frac{\alpha_{3 1}}{2}  \simeq 90^o, ~\nonumber \\ 
|\lan m_{\bt \bt }\ran |\simeq 0.032~{\rm eV}.~~~~~~~~~~~~~~~
\la{del}
\end{eqnarray}

 The phase and angles are given according to the standard parametrization \cite{partdata}. For different values for $\rho$ and even for  $\rho \rightarrow 0$ the mixing angles shown in (\ref{exa}, \ref{del}) are almost unaffected. For $\rho=0$ the largest mass is $m_2=0.0492~$eV.
In view of our simple approach
these results for the mixing angles and the mass squared differences are satisfactory. Nevertheless, in case the inverted hierarchy predicted here turns out to be established by experiment, a more detailed study of the model will be necessary. By integrating out heavy states renormalization group effects and the violation of unitarity of the mixing matrices must certainly be incorporated before any final judgement is possible.

\section{The scalar sector and the Higgs Boson.}\label{sec:4}

The embedding of the Standard Model into a larger group implies an extended Higgs structure formed by numerous scalar fields. This is difficult to deal with since only the information about the just discovered Higgs boson \cite{LHC} can be incorporated. Our aim is to construct in a phenomenological way examples of  tree-level potentials for the scalar fields and to calculate the corresponding boson mass spectrum.  The tree potential  has to be formed from $ SU(3)_L \times SU(3)_R$ invariants. 
 As mentioned in section 1 our input consists of vacuum expectation values only.  They determine the spontaneous symmetry breaking pattern and fix the position of the minimum of the potential. Clearly, the hierarchy problem is not solved this way but appears in a somewhat different light. The vevs, which are not understood anyhow,  have to be partly taken from experiment and partly to be postulated. They are fine-tuned with respect to radiative corrections.

The presence of Higgs fields with different properties and the necessity of combining  gauge group invariants in order to get non-zero masses leads to interesting properties of the obtained bosons. Some will drastically differ from the Standard Model Higgs-like states. We will discuss here two scenarios of interest. 

\subsection{The fermiophobe model.}
The potential responsible for the scalar particle spectrum is constructed from invariants of the fields in $H$  and $\tilde H$.  From these $36$ real fields   $21$  of them should become massive while leaving $15$ would-be-Goldstone particles massless. The remaining fields $H_A$ and $H_{A l}$ are supposed to have little influence on the scalar particle spectrum, at least not in the TeV region or below. Starting from a massless Lagrangian the individual invariants for $H$ and $\tilde H$ are 
\begin{eqnarray}
J_1 = (Tr[H^\dagger \cdot H])^2 , ~J_2  = 
Tr[ H^\dagger  \cdot H \cdot H^\dagger \cdot H] , ~~~ \nonumber \\
J_3 = (Tr[\tilde H^\dagger \cdot \tilde H])^2,~ J_4 = 
Tr[\tilde H^\dagger  \cdot \tilde H \cdot\tilde H^\dagger \cdot \tilde H] .~~~~~~~
\label{J}
\end{eqnarray}
with the vevs shown in (\ref {vevH}).  But in the following we will restrict the vevs $M_R$ and $M_3$ by $M_R^2 + M_3^2 = M_I^2$ and later take $M_3=0$.

Only $J_1$ and $J_3$ can be modified as in (\ref{1.1})  in order to have no linear terms after the appropriate shift $H \rightarrow \langle H \rangle + H $ and a similar shift for $\tilde H$. But the masses obtained from these two invariants and combinations of them are of order $M_I$, $M_R$. None of them are of order $v$ like the Higgs boson observed at the LHC. Thus one has either to use a different form or to add immediately new invariants which combine the fields of $H$ and $\tilde H$. 

\subsubsection{Potentials with logarithmic terms.}
Let us first  remain with the four important invariants (\ref {J}) but allow a logarithmic dependence on  $J/\langle J \rangle$ as in (\ref {1.2}).  Even though  a good justification for the form (\ref {1.2}) cannot be
given,  we can nevertheless use it for our phenomenological potentials. The reason is that by expanding the potential (after shifting the fields) in terms of the very large vevs occurring in our model and by neglecting inverse powers of these vevs, the so obtained effective potential has scalar fields up to the fourth power only. 

A naive Ansatz for the potential constructed from the four invariants shown above is 
 \begin{eqnarray}
\label{V0}
V_0 = \sum_{i=1}^{4}  ~\lambda_i ~\frac{J_i}{1+\log[\frac{J_i}{\lan J_i \ran}] }.
\end{eqnarray}
The second derivatives of each term in (\ref{V0}) (considered independently)  give  a mass matrix with a single non-zero eigenvalue $M_i^2$ that in the limit of a large value for $M_I$ ($M_R$) is dominated by $M_I$ ($M_R$):
 \begin{eqnarray}
\label{Vi}
M_i^2  \rightarrow  8~\lambda_i~ M_{I(R)}^2
\end{eqnarray}

If the differences  between $M_{I (R)}$ and the masses  generated remain finite for $M_{I (R)} \rightarrow \infty$ , i.e. $M_i^2  \rightarrow ~ M_{I(R)}^2 $, one  gets the interesting result 
 \begin{eqnarray}
\label{Vla}
 \lambda_1  = \lambda_2 =\lambda_3 =\lambda_4  = \frac{1}{8}.
\end{eqnarray}

The limit $M_i \rightarrow M_{I(R)}$  was speculatively assumed in \cite{BHiggs1}. The corresponding tree-level potential provided a prediction of the Higgs mass 
 $m_{Higgs}^2 \simeq \frac{v}{\sqrt{2}} \simeq 123 ~$GeV not far from the experimental value found later.  A similar ansatz for $V_0$ containing  only a single log function has been described in \cite{BTHiggs}. In the following we use $\lambda_1=\lambda_2=\lambda_3=\lambda_4=\frac{c_0}{8} $ allowing thereby for  a correction factor $c_0$ sightly different from one \cite{BHiggs2}.

From the second derivatives of
$\frac{1}{2}~V_0$ with respect to all $36$ fields at the point $H = \tilde H=0 $ of the shifted fields one gets the $36 \times 36$ mass matrix whose eigenvalues - shown here  for large $M_I$,  $M_R=M_I$  -  are
 \begin{eqnarray}
 \label{eigen} 
 \nonumber
m_1^2  =  \frac{c_0}{2}~(v_1^2 + b^2), ~~~~~m_2^2 = \frac{c_0}{2} (v_2^2 + b_3^2),~~~~~
m_3^2  =   2 ~c_0~ M_I^2,~~ ~~~  m_4^2 = 2 ~c_0~M_I^2 ,  \\
 ~~~~~~~~m_i^2  = 0  ~~~~~~~~ i = 5 ..............36~~~~~~~~~~~~~~
\end{eqnarray}
We note that this result can also be obtained by expanding $V_0$ (after the shift of fields) in terms of $M_I$ and neglecting  inverse powers of $M_I$.
In (\ref{eigen}) we have  $2$ scalars with low masses \cite{BHiggs2}. The first one can be identified with the Higgs boson found at the LHC. It is coupled to fermions and gauge bosons. Its mass is $m_{Higgs}^2 = \frac{c_0}{2}~ (v_1^2+b^2)$. The second boson is fermiophobic. It is not directly coupled to fermions, only to  gauge bosons.  Its properties and mass are sensitive to invariants not yet used.

$v_1$ and $v_2$ are strongly constrained by (\ref{v}). 
Taking   $b_j$, $f_j $ in (\ref{v}) to be small ($\approx m_b$)  one can write $v_1= v~\cos{\tilde\beta} $ and $v_2= v~ \sin{\tilde \beta} $.
$\tilde \beta =0$ implies $ g_t v = m_t$ ($g_t \simeq 1$)  as in the Standard model. Comparing then the value of the Higgs mass measured at the LHC with the expression for $m_1$ in 
(\ref{eigen}), $c_0$  is determined to be $\simeq 1.04 $ which is indeed close to $1$ as expected from the potential (\ref{V0}) with (\ref{Vla}).  An interesting but less likely case would be $\tilde\beta= \pi/4$ and therefore $m_1 = m_2$, i.e. a twin structure of the Higgs particle \cite{BTHiggs}. Its mass requires however the correction factor to be $c_0 \simeq 2$. Moreover, further contributions to the potential will in general remove this degeneracy as we will see below. 

The $32$ massless states in (\ref{eigen}) can be divided into $15$ massless Goldstone states and $17 $ additional states of mass zero. One massless state is due to the so far unbroken general phase transformation of $H$ and $\tilde H$. 
The remaining  $16 $ massless states are due to our provisional neglecting of invariants that connect the fields in $H$ with the fields in $\tilde H$. Without them $H$ or $\tilde H$ can be independently transformed by $SU(3)_L\times SU(3)_R$ matrices.  
                                                                                                                                                                                                                                          
There are a number of different invariants containing the fields of both multiplets $H$ and $\tilde H$ \cite{TriniPak}. The vevs proposed and  the parameters for the invariants have to be restricted to  guarantee that all scalar masses are positive and not in conflict with the data. We cannot perform this task in general since the  vevs and the allowed range of the dimensionless parameters depend on each other in a complicated way. Moreover,  the addition of a
new vev, or even a slight change of the ratio of two vevs,  can abruptly distort the spectrum. These properties provide 
 strong restrictions for the vevs and couplings that have not yet been explored.

Here we confine ourselves to the possible close connection between the  vevs of $H$ and $\tilde H$, namely  the correspondence 
\beqn
\label{MR}
M_R = M_I,~ ~M_3=0, ~~b_2 = 0~\text{ and}~~b_3~\text{of order}~ b
\eeqn
($M_R=M_I$, $b_2=0$ and  $b_3=-b$ correspond to a $\frac{\pi}{2} ~U_R$-spin rotation of the vev of $\tilde H$ with respect to the vev of $H$).

The potential to be added to $V_0$ needs at least $3$ new invariants of dimension $4$.  In order to have no linear terms in the shifted form of the potential one has to add  "induced" invariants of dimension $2$ and $3$ quite similar as in (\ref{1.1}). The coefficients of these latter invariants are not free but determined  by the vevs and couplings of the basic  invariants of dimension $4$ (and vanish for vanishing vevs).

The new invariants we take are
 \begin{eqnarray}
\label{J2}
\hs{0cm}  J_5=  Tr [H^{\dagger}\cdot \tilde H \cdot \tilde H^{\dagger} \cdot H],  ~~~
&J_6& = Tr [H^{\dagger}\cdot H \cdot \tilde H^{\dagger} \cdot  \tilde H],~ ~~
J_7 =  Tr  [H^{\dagger}\cdot \tilde H \cdot  H^{\dagger} \cdot \tilde H] +       
Tr  [\tilde H^{\dagger}\cdot H \cdot \tilde H^{\dagger} \cdot H], ~~~~~ \nonumber   \\
J_8 = Tr [ H^{\dagger}  H],~~~~~~&J_9& =Tr [\tilde H^{\dagger} \tilde H], ~~~~~~~~~
J_{10} = \det{H} +\det {H^{\dagger}},~~~~~~J_{11} = \det{\tilde H} +\det {\tilde H^{\dagger}}.
\end{eqnarray}
Here the second line contains the  "induced" invariants with dimension $2$ and $3$.
The potential reads
\begin{eqnarray}
 \label{VS} 
 &V &= V_0  +  V_S,   \nonumber \\
&V_S &=   r_1 J_1 +r_2 J_2 + r_3 J_3 + r_4 J_4  +  r_5 J_5 + r_6 J_6 + r_7 J_7 + \mu_1^2 J_8 + \mu_3^2 J_9 + \mu_{d1} J_{10}+\mu_{d2} J_{11}.
\end{eqnarray}

The coefficients $r_1...r_4$ are required to be very small 
since the corresponding invariants appear already in $V_0$. 
The first ten invariants $J_1$ to $J_{10}$  do not change under the $P_G$ transformation   $ H \rightarrow H $ and $\tilde H \rightarrow - \tilde H $, while  $J_{11}$  changes sign. $J_{10}$ and  $J_{11}$  break the invariance under a common phase transformation of the fields. The total potential $ V $ should now provide non-zero masses for all fields except the Goldstone ones.

After requiring the vanishing of all first derivatives of $V$ at the proposed minimum, $\mu_1^2$, $\mu_3^2$, $\mu_{d1}$ and $\mu_{d2}$ are determined  by our vevs  and the dimensionless couplings.  For simplicity, and also to avoid near-by negative eigenvalues of the mass matrix,  we set in the following   $r_5 = - 2 r_1-2 r_2$ and in addition $r_2=r_1$. 
Let us consider the cases a)~ $\tilde \beta = 0$~ i.e. $v_1 = v, v_2=0$ with $c_0 \simeq 1$~~ and ~~ b)~ $\tilde \beta = \frac{\pi}{4}$ ~i.e. $ v_1= v_2 = \frac{v}{\sqrt{2}}$   with $c_0 \simeq 2$. 

\subparagraph{Case a)}

 By setting $v_2=0$ the minimum condition  for $V$  fixes $r_6$, $r_7$, $\mu_1^2$, $\mu_3^2 $,  $\mu_{d1}$, $\mu_{d2}$  in terms of $r_1$, $r_3$ and $r_4$.  In this example we take $M_I = M_R = 10^{13}~$GeV, $b= 2.85~$GeV and $b_3= 1~$GeV. For $r_4=0$  and almost 
  independent of $r_1, r_3$, the Higgs boson obtained in (\ref{eigen}) appears again,  this time together with an acceptable mass spectrum for the other $20$ bosons. For $r_1 \gtrsim v^2/M_I^2$ it
 is the lowest scalar state. Its mass is~$\sqrt{c_0}\cdot123~$GeV.  
 
 This Higgs field is to $99.9 \%$ composed  of the field $^1 H_1$ with an admixture of $^2 H_2$ in the ratio $ b / v =0.016$.  Only a tiny $0.57 \%$ admixture of the fermiophobic field $^2 \tilde H_3$ can be noticed. The coupling to $t$ and  $b$ quarks (the latter due to the appearance of
the Higgs eigenstate in the field $^2H_2$ with the factor $b/v$) is identical to the one of the Standard Model. On the other hand, the Higgs self-coupling constants for  $H_{Higgs}^3$ and 
  $H_{Higgs}^4 $  are twice as large as in the Standard Model. This increase is due to the logarithmic terms occurring in $V_0$.
  For large values of  $M_I$ this potential gets a polynomial form generating these couplings. These self-coupling constants and the tiny admixture with the fermiophobic field (vanishing for $b_3 \rightarrow 0$) are the only deviations from Standard Model properties. Their determination can be used to reject or support this potential. 
  
  The properties of the higher states, however, are strikingly different from conventional Higgs particles.  
 For  $ r_1 = 6 v^2/M_I^2$, for example, the next states are $4$ almost degenerate scalars at $\simeq 852~$GeV.  It is a complex $SU(2)_L$  doublet $\tilde{^i H_1}$ ($i=1, 2$) and thus purely fermiophobic. The next $4$ states at $1205~$GeV are again degenerate. They are the members 
 of the $SU(2)_L$ doublet $\tilde{^i H_2} - ^i H_3 $ with an equal amount of normal and fermiophobic components.
 Taking $r_1$ to be very small (f.i. $r_1= 0.01 v^2/M_I^2$) the pure fermiophobic states are shifted below the Higgs at $125~$GeV. 
 
  Interestingly, it is also possible to have a solution with $\mu_1^2=\mu_3^2 =0$. Here  $r_3$ and $r_4$ have to have special values to achieve this result.   $J_{11}$ is then the only term with dimension different from $4$ appearing in the potential (\ref{VS}). Its coefficient $\mu_{d1}$ is fixed by the vevs and dimensionless couplings of the invariants with dimension $4$.
   $\mu_{d2}$ remains $0$. Relevant mass spectra and field compositions are presented in the appendix. 
   
\subparagraph{Case b)}

For  $ v_1= v_2 = \frac{v}{\sqrt{2}}$ and $c_0 \simeq 2$ we take as before  $M_I = M_R = 10^{13}~$GeV, $b= 2.85~$GeV  and this time $b_3= \pm b$. We also set $r_5 = - 2 r_1 -2 r_2 $, keep $r_3 = 0$ and  choose as an example $r_2=10^{-5},~ r_6=1$. This leaves the parameter $r_1$.   For values 
$r_1 \gtrsim  2 v^2/M_I^2$ the $125$~GeV Higgs is the lowest state  separated by a sizable mass gap from  scalars with higher masses. 

However, this particle is now different from a normal Higgs: it is an equal admixture of the fields $H$ and $\tilde H$ (see table 3 in the appendix).  But even though  this  structure is quite different from a normal Higgs and the top coupling $m_t/v_1$ to the field $^1H_1$ is larger by the factor $ {\sqrt{2}}$,  the couplings of this boson to fermions remain equal to the Standard Model couplings. The reason is that only the $1/\sqrt{2}$ part of the field $^1H_1$ forms the Higgs eigenstate and thus compensates these effects. The calculation shows that the invariant $J_{11}$ has now a non-vanishing coefficient. Thus, the chosen vevs together with the dimensionless couplings enforce a breaking of the $P_G$ symmetry. For the $r_1$ value used in Table III its magnitude is $\mu_{d2}=1.57 \cdot 10^{-6}$~GeV. To force it to be zero would imply the existence of several massless bosons. Due to the logarithmic parts in the potential the Higgs self-couplings are, as in the examples given before, a factor $2$ larger than in the Standard Model. All higher states differ drastically from the Standard Higgs boson because of their very different field components.

For a special choice of the parameter $r_1$  ($ r_1\simeq 10^{-3} v^2/M_I^2$)  and $b_3=+b$ a twin structure for the Higgs (mass degenerate twins) can be achieved. One member has the orthogonal admixture of normal and fermiophobic scalars as compared to the second member. Our model allows this degeneracy but  does not  favor it. To exclude or verify a mass degeneracy the method of ref \cite{Gross} could be applied.  
This twin is the lowest state, but by going further down with $r_1$ other states can lie below.   

\subsubsection{Potential without logarithmic terms.}
We now turn to an example for a potential without the logarithmic terms  occurring in $V_0$. We simply leave out $V_0$ in (\ref{VS}).  As before the minimum condition determines the coefficients of the invariants of dimension $2$ and $3$. We take $M_I=M_R=10^{13}~$GeV, $b=2.85~$GeV, $b_3 = 1~$GeV and $v_1 =v~\cos{\tilde \beta},~ v_2=v ~\sin{\tilde \beta}$. This time the particle corresponding to the Higgs boson found at the LHC does no longer stick out within large ranges of the  $r$ parameters as in the cases treated above. Now these parameters have to be carefully fine-tuned to obtain the relevant field composition with the correct mass. They also depend strongly on the value taken for  $\tilde \beta$.
 Still $r_3$ can be set  equal to $1$ since it affects mainly the highest masses.  Taking again $r_1=r_2,~ r_5= -4 r_1$ the  
 boson corresponding to the Higgs found at the LHC can then be fixed by the parameters $r_1$ and $r_4$. They need fine-tuning to avoid negative mass squared values for one or more of the other $20$ states.  For $\tilde \beta = 0$ the  Higgs field is composed of the field $ ^1H_1 + \frac{b}{v} ~^2H_2$ as before. Its largest fermiophobic component $^2 \tilde H_3$  is again only $0.57 \%$ of the Higgs field. Its coupling to top and $b$ quarks and now also its self-coupling constants are practically identical to the Standard Model couplings.Thus, this boson differs only minimally from the Standard Model Higgs boson.  Unfortunately, all higher states are far away at about $10^{10} -10^{13}~$GeV as a consequence of the high value we took for $M_I$. The composition of the Higgs field for $3$ different values of $\tilde \beta$ and the corresponding values of $r_1$ and $r_4$ are shown in Table III in the appendix.

\subsection{The flavophile model}
Here we set $\tilde H = 0 $ and thus have no fermiophobic components in the scalar particle spectrum. Instead, the potential is constructed from the fields occurring in $H$ and $H_A$. This is another way to obtain a boson spectrum with no vanishing masses except the $15$ would-be-Goldstone bosons. Unavoidably, it will  lead to flavor-changing contributions. The fields from $H_{A l}$ are supposed either not to be relevant for low-lying states and/or to have their own additional potential not involving the other fields. This is required because the existence of  a state coupled to leptons  having simultaneously flavor conserving and flavor changing components  would hardly be consistent with the known strong limits  \cite{mu e}  on the decay $ \mu \rightarrow  e ~ \gamma $ .

For the scalar potential  we take the same approach as in the fermiophobe model and simply replace $\tilde H$ by $H_A$. Now we have to set $v_1\simeq v$,~ $v_2 \rightarrow ~^1 f_1\rightarrow 0$ , $b_2 \rightarrow~  ^2 f_2 $, $b_3 \rightarrow~ ^2 f_3 $,  $M_R \rightarrow~ ^3 f_2$ and $M_3 \rightarrow ~^3f_3$.  
 As in (\ref{MR}) we use again $^3 f_2 = M_I$, $^3 f_3 = 0$, $^2f_2=0$ and $^2f_3 =1$~GeV. 
 
Apart from  the completely  different interpretation one can copy the results from the fermiophobe model. In analogy to case a) above there is a large range of values for the parameters $r_1$ and $r_3 $ for which the state at $\simeq 125$~GeV coincides with the low mass Higgs boson in (\ref{eigen}). It is barely affected by the heavier bosons. The additional field component  $ (^2(H_A)_3)$ that can lead to flavor-changing transitions via the matrix A is again only $\simeq 0.57 \%$  of the field $^1H_1$. This boson can hardly be distinguished from the Standard Model Higgs. 
Nevertheless, a careful analyses of the decay amplitudes resulting from flavor-changing components of the Higgs field, like the one performed in  \cite{Kopp}, would be highly desirable.  For almost all other states our model predicts field components from $ H_A$ with similar strength as the fields from $H$. Such neutral and charged fields could also lie below the Higgs.  If such states exist one could look for their decays to two quarks of different flavors, for instance to a jet with a leading bottom quark and a jet with a leading strange quark.   An intensive search for such decays as well as for low-energy processes induced by virtual boson exchange (in analogy to penguin-type processes) is suggested.  Flavor-violating Higgs bosons would be of significance for an increased understanding of the CKM matrix.    With regard to  potentials  without the $V_0$ part of the potential in (\ref{VS}) one can see that here the obtained boson also coincides very nearly with the Standard Model Higgs while all other states differ considerably from normal Higgs fields.  
 \vs{0.0cm}
\section{Summary}\label{summary}

In this work we considered the generalization of the Standard Model to the trinification group $SU(3)_L \times SU(3)_R \times SU(3)_C$  augmented by the  generation symmetry $SO(3)_G$ under which all fermions are $3$-vectors in generation space. In our phenomenological approach essential use is made of the vacuum expectation values of  scalar Higgs fields. They provide the spontaneous symmetry breaking down to the Standard Model and finally to $U(1)_e$.  As in \cite{BZ2,BZ3} an effective Yukawa interaction is proposed that,  beside flavor singlet Higgs fields, contains two flavor (generation) matrices $G$ and $A$. $G$ determines the mass hierarchy of all fermions and $A$ all mixings. The difference between the up quark spectrum and the spectrum of down quarks as well as the structure of the CKM matrix are related to the mixing of fermions with heavy states present in the group representation. 
The form of the neutrino mass matrix is determined by this mixing as well. Our model leads to an inverted neutrino hierarchy.  With the measured atmospheric mass squared difference as input and having only one fit parameter, a satisfactory result for the experimentally observed solar neutrino mass difference and the neutrino mixing pattern is achieved. 

Concerning the Higgs sector only simple examples could be dealt with due to the many scalar fields in the representation of
 the trinification group.  Phenomenological tree-level potentials have been constructed giving mass to all fields apart from the $15$ would-be Goldstone particles. Due to the combination of different matrix fields required to obtain finite masses,  the mass eigenstates have either fermiophobic components or parts that induce flavor changing processes. A notable exception is the  Higgs-like state at $125$~GeV which appears independent of a wide range of parameter values. It has the same gauge and fermion couplings as the Standard Model Higgs and barely differs from it in this respect. A strong difference only occurs in the self-coupling constants:  The logarithmic part of the potential advocated here enforce them to be larger by a factor $2$ compared to the Standard Model.  
 
 In contrast to this particle all other scalar states strongly differ from usual Higgs-like  bosons. These new bosons, if existing at all, have interesting properties and would allow the study of exciting new processes. 
 \vs{0.2cm}
 
{\bf Acknowledgments}

It is a pleasure to thank David Lopez-Val and Tilman Plehn for many stimulating discussions.

\section{Appendix}
\subsection{Fermion masses at the scale $m_Z.$}
\vs{-0.7cm}
\begin{table}[hbt]
\caption{\label{table:fermionmasses}
Fermion masses at scale $\mu=m_Z$ in the $\overline{\rm MS}$ scheme calculated in \cite{Jamin} using the masses from \cite{PDG}
except for the $u-$, $d-$, and $s-$quark masses taken from \cite{FLAG} at $\mu=2$ GeV. }
\vs{0.3cm}
\centering
\begin{tabular}{|c|c|c|}
\hline
$m_u=1.24 \pm 0.06$ MeV   &   $m_c=624\pm 14$   MeV     &    $m_t=171.55\pm 0.90$ GeV \\
$m_d=2.69 \pm 0.09$ MeV & $m_s=53.8\pm 1.4$ MeV         &    $m_b=2.85\pm 0.023$  GeV \\
$m_e=  0.510~$ MeV                & $m_\mu = 105.4~$MeV                       & $m_\tau=1772.5~$MeV  
\\
\hline
\end{tabular}
\end{table}

\subsection{Scalar masses for $v_1= v$, $v_2=0.$}

We take $M_I=10^{13}$~GeV, $v=174~$GeV, $b=2.85$~GeV, $b_3= 1$~GeV, $c_0 =1.04$ and use for illustration $r_1=r_2= 6 v^2/M_I^2$,  $r_3=1$, $r_4=0$, $r_5= -4 r_1$.
The minimum conditions fix then the remaining $r$ coefficients: \\
$r_{6}=1.54\times 10^{-17}$, 
$r_{7}= - 2.71\times 10^{-18}$, $\mu_1^2= - 2.22\times 10^{-20}$ GeV$^2$,  $\mu_3^2=-2.00\times 10^{26}$ GeV$^2$, $\mu_{d_1}=-2.22\times 10^{-6}$ GeV, 
$\mu_{d_2}=0.~$The corresponding spectrum is shown in Table II.

\begin{table}[hbt]
\caption{\label{table:scalar masses}
Scalar mass spectrum $v_1=v$,~$v_2=0$.}
\centering
\begin{tabular}{|c|c|}
\hline
masses [GeV]   &  field composition (i=1,2)    \\
\hline
   &    \\
$\sqrt{\frac{c_0}{2} } ~v=125.5$  \qquad    & \qquad      ${\rm Re}[ 0.999\;{}^1H_1 +0.0164\;{}^2H_2+0.0057\;{}^2\tilde H_3]$  \qquad\\

852 (4 states)           &    ${}^i\tilde H_1$  \\

1205 (4 states)            &   $\frac{1}{\sqrt2}\left( -  {}^i H_3 + {}^i\tilde H_2 \right)$  \\

31667 (4 states)           &   $\simeq\frac{1}{\sqrt2}\left( {}^i H_2 + {}^i\tilde H_3 \right)$  \\

44766                     &   ${\rm Re}\left[\frac{1}{\sqrt2}\left( {}^3 H_2 + {}^3\tilde H_3 \right)\right]$  \\

45680 (4 states)           &   $\simeq\frac{1}{\sqrt2}\left( {}^i H_2 - {}^i\tilde H_3 \right)$  \\

64591                      &   ${\rm Im}\left[\frac{1}{\sqrt2}\left( {}^3 H_2 - {}^3\tilde H_3 \right)\right]$  \\

$1.4\times 10^{13}$         &   ${\rm Re}\left[ {}^3 H_3\right]$  \\

$2.5\times 10^{13}$         &   ${\rm Re}\left[ {}^3 \tilde H_2\right]$  \\

\hline
\end{tabular}

\end{table}

A very similar mass spectrum like the one in the Table II is obtained for $r_1= 6 v^2/M_I^2$ with the restriction $\mu_1^2=\mu_3^2=0.$  For this solution $r_3$ and $r_4$ have to have the special values  
$r_{3}=1.48\times 10^{-20}$, 
$r_{4}=-1.11\times 10^{-20}$ while  $r_6$, $r_7$, $\mu_{d_1}$, $\mu_{d_2}$ come out as above.

\subsection{Mass and composition of the 2 lowest states for the case of $v_1=v_2=\frac{v}{\sqrt{2}}.$} 

We take the same input values as in case b) described in the text and choose $r_1 = 6~ v^2/ M_I^2$ and $b_3=b=2.85$~GeV.
\begin{table}[hbt]
\caption{\label{table:Higgs} Higgs boson at $125$~GeV and the next higher state at $615$~GeV for $v_1=v_2$. The coefficients resulting from the \\
\vs{0.2cm}
minimum condition are: $r_7= - 0.50$,~ $\mu_1^2=-5.5 \cdot 10^{-27} ~$GeV$^2$, 
~$\mu_3^2=3.0 \cdot 10^5$~GeV$^2$,
~$\mu_{d1}= -\mu_{d2}=  -1.57 \cdot 10^{-6}~$GeV.
\\}.
\centering
\begin{tabular}{|c|c|}
\hline
mass [GeV]   &  field composition    \\
\hline
   &    \\
$\sqrt{c_0} \frac{v_1}{\sqrt{2}}=125$  \qquad  & \qquad     ${\rm Re}[ 0.7069\;{}^1H_1 + 0.7069\;{}^1\tilde H_1 +0.0161\;{}^2H_2+0.0161\;{}^2\tilde H_3]$  \qquad\\

$615$     & ${\rm Re}[ 0.7071\;{}^1H_1 - 0.7071\;{} \tilde{^1H_1} ]$ \\

\hline
\end{tabular}\\
\end{table}

\subsection{Higgs scalars depending on $\tilde \beta$. Potential without logarithmic terms.}

\begin{table}[hbt]
\caption{\label{table:Higgs scalars}
Composition of Higgs scalars ($m_{Higgs}=125~$GeV)  for $V_0=0$, $V=V_S$. 
 $v_1=v \cos{\tilde \beta}$ , $v_2=v \sin{\tilde \beta}$, 
$r_1=r_2$, $r_5=-4 r_1$,  $r_3 =1$. We use $r_4=3.01\cdot 10^{-4}$  for all $3$  $\tilde \beta $ values and fit $r_1$ for each case. \\}. 
\centering
\begin{tabular}{|c|c|}
\hline
$125~$GeV, $\tilde \beta$,  $r_1$ &  field composition    \\
\hline
   &    \\
$\tilde \beta =0$,
$r_1=6.71\cdot 10^{-2}$  \qquad  & \qquad      ${\rm Re}[ 0.999\;{}^1H_1 +0.016\;{}^2H_2+0.0057\;{}^2\tilde H_3]$  \qquad\\

$\tilde\beta = \frac{\pi}{8} $, $r_1=1.82\cdot 10^{-4}$      & ${\rm Re}[ 0.9237\;{}^1H_1 + 0.3826\;{} \tilde{^1H_1} +0.0161\;{}^2H_2+0.00575\;{}^2\tilde H_3]$ \\

$\tilde \beta =\frac{\pi}{4}$, $ r_1=1.552\cdot 10^{-4}$        &   ${\rm Re}[ 0.7070\;{}^1H_1 + 0.7070\;{} \tilde{^1H_1} +0.0161\;{}^2H_2+0.00575\;{}^2\tilde H_3]$\\

\hline
\end{tabular}\\

\end{table}

\newpage
\bibliographystyle{unsrt}

\end{document}